\documentclass[]{jfm}

\usepackage{graphicx}
\usepackage{epstopdf,epsfig}
\usepackage{newtxtext}
\usepackage{newtxmath}
\usepackage{natbib}

\newcommand{\RomanNumeralCaps}[1]
\linenumbers

\usepackage{bm}%
\usepackage[colorlinks,
            linkcolor=blue,
            anchorcolor=blue,
            citecolor=blue
            ]{hyperref}
\usepackage{subfigure}
\usepackage{multirow}
\usepackage{natbib}
\usepackage{float}
\usepackage{makecell}
\usepackage{amsmath}

\usepackage{color}

\shorttitle{Consistency between AEM and IOIM}
\shortauthor{C. Cheng \&  L. Fu}

\title{Consistency between the attached-eddy model and the
inner-outer interaction model: a study of streamwise wall-shear stress fluctuations in a turbulent channel flow}

\author{
Cheng Cheng\aff{1}, 
Lin Fu\aff{1}$^,$\aff{2}$^,$\aff{3}$^,$\aff{4}$^,$\corresp{\email{linfu@ust.hk}}
}

\affiliation{
\aff{1}Department of Mechanical and Aerospace Engineering, The Hong Kong University of Science and Technology, Clear Water Bay, Kowloon, Hong Kong
\aff{2}Department of Mathematics, The Hong Kong University of Science and Technology, Clear Water Bay, Kowloon, Hong Kong
\aff{3} Center for Ocean Research in Hong Kong and Macau (CORE), The Hong Kong University of Science and Technology, Clear Water Bay,
Kowloon, Hong Kong
\aff{4}Shenzhen Research Institute, The Hong Kong University of Science and Technology, Shenzhen, China
}

\begin{document}
\maketitle

\begin{abstract}
The inner-outer interaction model (Marusic, Mathis \& Hutchins, Science, vol. 329, 2010, 193-196) and the attached-eddy model (Townsend, Cambridge University Press, 1976) are two fundamental models describing the multi-scale turbulence interactions and the organization of energy-containing motions in the logarithmic region of high-Reynolds number wall-bounded turbulence, respectively. In this paper, by coupling the additive description with the attached-eddy model, the generation process of streamwise wall-shear fluctuations, resulting from wall-attached eddies, is portrayed. Then, by resorting to the inner-outer interaction model, the streamwise wall-shear stress fluctuations generated by attached eddies in a turbulent channel flow are isolated. Direct comparison between the statistics from these two models demonstrates that they are consistent to and complement each other. Meanwhile, we further show that the superpositions of attached eddies follow an additive process strictly by verifying the validity of the strong and extended self similarity. Moreover, we propose a Gaussian model to characterize the instantaneous distribution of streamwise wall-shear stress, resulting from the attached-eddy superpositions. These findings are important for developing an advanced reduced-order wall model. 

\end{abstract}

\begin{keywords}
\end{keywords}

{\bf MSC Codes }  {\it(Optional)} Please enter your MSC Codes here

\section{Introduction}

Wall-shear stress fluctuation is a crucial physical quantity in wall-bounded turbulence, as it is of importance for noise radiation, structural vibration, drag generation, and wall heat transfer, among others \citep{Diaz-Daniel2017,Cheng2020a}. In the past two decades, ample evidence has shown that the root mean squared value of streamwise wall-shear stress fluctuations ($\tau_{x,rms}^{'}$) is sensitive to the flow Reynolds number \citep{Abe2004,Schlatter2010,Yang2017,Guerrero2020}. It indicates that large-scale energy-containing eddies populating the logarithmic and outer regions in high-Reynolds-number wall turbulence have non-negligible influences on the near-wall turbulence dynamics, and thus the wall friction \citep{Giovanetti2016,Li2019a}. 

Till now, several models have been proposed on the organization of motions in logarithmic and outer  regions and their interactions with the near-wall dynamics. \cite{Marusic2010} have established that superposition and modulation are the two basic mechanisms that large-scale motions (LSM) and very-large-scale motions (VLSM) exert influences on the near-wall turbulence. The former refers to the footprints of LSMs and VLSMs on the near-wall turbulence, while the latter indicates the intensity amplification or attenuation of near-wall small-scale turbulence by the outer motions. \cite{Mathis2013} extended the model to interpret the generation of wall-shear stress fluctuations in high-Reynolds number flows. They emphasized that superposition and modulation are still two essential factors. This inner-outer interaction model (IOIM) has also been successfully developed to predict the near-wall velocity fluctuations with data inputs from the log layer \citep{Marusic2010,Baars2016,Wang2021}.

On the other hand, the most elegant conceptual model describing the motions in logarithmic region is the attached-eddy model (AEM) \citep{Townsend1976,Perry1982}. It conjectures that the logarithmic region is occupied by an array of self-similar energy-containing motions (or eddies) with their roots attached to the near-wall region. Extensive validations support the existence of attached eddies in high-Reynolds number turbulence, such as the logarithmic decaying of streamwise velocity fluctuation intensities \citep{Meneveau2013}, as originally predicted by \cite{Townsend1976}. The reader is referred to a recent review work by \cite{Marusic2019} for more details. Given the existence of wall-attached energy-containing motions in the logarithmic region, it would be quite natural to hypothesize that the near-wall part of these motions would affect the generation of the wall-shear fluctuations to some extent, maybe, via the superposition and modulation mechanisms. However, some fundamental questions may be raised, e.g., whether the IOIM and AEM are consistent with each other? There's a possibility that the superposition component of $\tau_x'$ decomposed by the IOIM in physical space can not fully follow the predictions made by the AEM quantitatively. If yes, whether these two models
can shed light on the mechanism of wall-shear fluctuation generation and be indicative for modeling approaches?

Previous study \citep{Yang2017} verified that the generation of wall-shear stress fluctuations can be interpreted as the outcomes of the momentum cascade across momentum-carried eddies of different scales, and modeled by an additive process. Here, we first aim to couple the additive description with the AEM to portray the generation process of streamwise wall-shear fluctuations, resulting from wall-attached eddies. Two scaling laws describing their intensities and the linkages with the characteristic scales of attached eddies can be derived (the characteristic scales of attached eddies are their wall-normal heights according to AEM \citep{Townsend1976}). Then, we intend to isolate the streamwise wall-shear stress fluctuations generated by attached eddies in a turbulent channel flow at $Re_{\tau}=2003$ ($Re_{\tau}=hu_{\tau}/\nu$, $h$ denotes the channel half-height, $u_{\tau}$ the wall friction velocity and $\nu$ the kinematic viscosity) by resorting to the IOIM \citep{Marusic2010,Baars2016}. Here, IOIM is employed as a tool to estimate the streamwise wall-shear fluctuations generated by attached eddies. The statistics from IOIM can be processed to verify the scaling laws deduced by AEM, so as to demonstrate their consistency. Moreover, a simple algebraic model describing the instantaneous distributions of the streamwise wall-shear stress fluctuations generated by attached eddies will be proposed.

\section{Streamwise wall-shear stress fluctuations generated by attached eddies}
According to \cite{Mandelbrot1974} and \cite{Yang2017}, the generation of streamwise wall-stress fluctuations can be modeled as an additive process within multifractal formalism, which takes the form of
\begin{equation}
\tau_{x}^{'+}=\sum_{i=1}^{n}a_i,
\end{equation}
where $a_i$ are random addends, representing an increment in $\tau_x^{'+}$ due to eddies with wall-normal height $h/2^i$, and superscript $+$ denotes the normalization with wall units. Here, we intend to isolate the contributions from the eddies populating logarithmic region ($\tau_{x,o}^{'+}$) and link to their wall-normal positions $y$. $\tau_{x,o}^{'+}$ can be expressed as
\begin{equation}\label{Gaa}
\tau_{x,o}^{'+}=\sum_{i=n_s}^{n_o}a_i,
\end{equation}
where $n_s$ and $n_o$ represent the additives that correspond to the eddies with the wall-normal height at $y_s$ and $y_o$, respectively. Here, $y_s$ is the lower bound of logarithmic region, and  generally believed to be  $80 \leq y_s^+\leq 100$ \citep{Jimenez2018,Baars2020a}; $y_o$ is the outer reference height. It can be found that $1\textless n_s \textless n_o \textless n$.
 The addends $a_i$ are assumed to be identically and independently distributed (i.i.d) and equal to $a$. The number of the addends should be proportional to
\begin{equation}
n_o-n_s+1\sim \int_{y_s}^{y_o} p(y) dy \sim \int_{y_s}^{y_o} \frac{1}{y}   dy\sim \ln(\frac{y_o}{y_s}),
\end{equation}
where $p(y)$ is the eddy population density, which is proportional to $1/y$ according to AEM \citep{Townsend1976,Perry1982}. A momentum generation function $\left\langle exp(q\tau_{x,o}^{'+})\right\rangle$, where $<>$ represents the averaging in the temporal and spatially homogeneous directions, is defined to scrutinize the scaling behavior of $\tau_{x,o}^{'+}$ \citep{Yang2016a}.
$\left\langle exp(q\tau_{x,o}^{'+})\right\rangle$ can be evaluated as 
\begin{equation}\label{SSS}
\left\langle exp(q\tau_{x,o}^{'+})\right\rangle=\left\langle exp(qa)\right\rangle^{n_o-n_s+1}\sim (\frac{y_o}{y_s})^{s(q)},
\end{equation}
where $q$ is a real number, $s(q)=C_1\ln\left\langle exp(qa)\right\rangle$ is called anomalous exponent, $C_1$ is a constant. Eq.~(\ref{SSS}) is called strong self similarity (SSS).
If $a$ is a Gaussian variable, the anomalous exponent can be recast as 
\begin{equation} \label{Ga}
s(q)=C_2q^2,
\end{equation}
where $C_2$ is another constant. On the other hand, an extended self-similarity (ESS) is defined to describe the relationship between $\left\langle exp(q\tau_{x,o}^{'+})\right\rangle$ and $\left\langle exp(q_0\tau_{x,o}^{'+})\right\rangle$ (fixed $q_0$) \citep{Benzi1993}, i.e.,
\begin{equation}\label{equ:M5}
\begin{aligned}
\left\langle exp(q\tau_{x,o}^{'+})\right\rangle = \left\langle exp(q_0\tau_{x,o}^{'+})\right\rangle^{\xi(q,q_0)},
\end{aligned}
\end{equation}
where $\xi(q,q_0)$ is a function of $q$ (fixed $q_0$). Note that ESS does not strictly rely on  i.i.d of the addends, but the additive process Eq.~(\ref{Gaa}). 

\section{DNS database and scale decomposition method}
The direct numerical simulation (DNS) database used in the present study is an incompressible turbulent channel flow at $Re_{\tau}=2003$, which has been extensively validated by previous studies \citep{Hoyas2006,Jimenez2008}.
The decomposition of $\tau'_x$ is based on the IOIM first proposed by \cite{Marusic2010}. \cite{Baars2016} modified the computational process by introducing spectral stochastic estimation to avoid artificial scale decomposition. In this work, the modified version of IOIM is adopted to investigate the multi-scale characteristics of  $\tau'_x$.
It can be expressed as
\begin{equation}
u_{p}^{+}\left(y^{+}\right)=\underbrace{u^{*}\left(y^{+}\right)\left\{1+\Gamma_{u u} u_{L}^{+}\left(y^{+}\right)\right\}}_{u_{s}^{+}}+u_{L}^{+}\left(y^{+}\right),
\end{equation}
where $u_{p}^{+}$ denotes the predicted near-wall streamwise velocity fluctuation, $u^{*}$ denotes the universal velocity signal without large-scale impact,  $u_L^{+}$ is the superposition component,  $\Gamma_{u u}$ is the amplitude-modulation coefficient, and $u_{s}^{+}$ denotes the amplitude modulation of the universal signal $u^{*}$. $u_{L}^{+}$ is obtained by spectral stochastic estimation of the streamwise velocity fluctuation at the logarithmic region $y_o^+$, namely,
\begin{equation}
u_{L}^{+}\left(x^{+}, y^{+}, z^{+}\right)=F_{x}^{-1}\left\{H_{L}\left(\lambda_{x}^{+}, y^{+}\right) F_{x}\left[u_{o}^{+}\left(x^{+}, y_{o}^{+}, z^{+}\right)\right]\right\}
\end{equation}
where $u_{o}^{+}$ is the streamwise velocity fluctuation at $y_o^+$ in the logarithmic region, and, $F_x$ and $F_x^{-1}$ denote FFT and inverse FFT in the streamwise direction, respectively. $H_L$ is the transfer kernel, which evaluates the correlation between $u^+(y^+)$ and $u_{o}^{+}(y_o^+)$ at a given length scale $\lambda_{x}^{+}$, and can be calculated as
\begin{equation}\label{HL}
H_{L}\left(\lambda_{x}^{+}, y^{+}\right)=\frac{\left\langle\hat{u}\left(\lambda_{x}^{+}, y^{+}, z^{+}\right) \overline{\hat{u}_o}\left(\lambda_{x}^{+}, y_{o}^{+}, z^{+}\right)\right\rangle}{\left\langle\hat{u}_o\left(\lambda_{x}^{+}, y_{o}^{+}, z^{+}\right) \overline{\hat{u}_o}\left(\lambda_{x}^{+}, y_{o}^{+}, z^{+}\right)\right\rangle},
\end{equation}
where $\hat{u}$ is the Fourier
coefficient of $u$, and $\overline{\hat{u}}$ is the complex conjugate of $\hat{u}$.

\begin{figure} 
	\centering 
	\subfigure{ 
		\label{fig:AEH:a} 
		\includegraphics[width=3.5in]{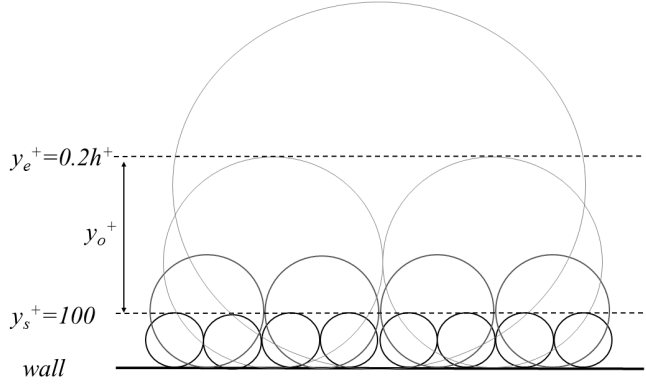} 
	}
	\caption{A schematic of the attached-eddy model \citep{Hwang2015}. Each circle represents an individual attached eddy. $y_s^+$ and $y_e^+$ are the lower and upper bound of the logarithmic region, respectively. $y_0^+$ is the outer reference height, and varies from $y_s^+$ to $y_e^+$.}
	\label{fig:AEH} 
\end{figure}

In this work, we mainly pay attention to the quantity, $\tau'_x$ generated by the attached eddies. Thus, the predicted position $y^+$ is fixed at $y^+=0.3$, and the outer reference height $y_o^+$ varies from $100$ (namely $y_s^+$) to $0.2h^+$ (denoted as $y_e^+$), i.e., the upper boundary of logarithmic region \citep{Jimenez2018}. We have checked that as long as the predicted position is around $y^+\le O(1)$, the results presented below are insensitive to the choice of specific $y^+$. Once $u_L^+$ is obtained, the superposition component of $\tau_x^{'+}$ can be calculated by definition (i.e., $\frac{ \partial u_L^{'+}} { \partial y^+}$ at the wall) and denoted as $\tau_{x,L}^{'+}(y_o^+)$. According to the hierarchical attached eddies in high-Reynolds number wall turbulence (see Fig.~\ref{fig:AEH}), $\tau_{x,L}^{'+}(y_o^+)$ represents the superposition contributed from the wall-coherent motions with their height larger than $y_o^+$. Thus, the difference value $\tau_{x,L}^{'+}(y_s^+)-\tau_{x,L}^{'+}(y_o^+)$ can be interpreted as the superposition contribution generated by the wall-coherent eddies with their wall-normal heights within $y_s^+$ and $y_o^+$, i.e., $\tau_{x,o}^{'+}$ in Eq.~(\ref{Gaa}). Considering that $y_s^+$ is the lower bound of the logarithmic region, the increase of $y_o^+$ corresponds to the enlargement of the addends in
the additive description (see Eq.~(\ref{Gaa})). In this way, the connection between AEM and IOIM are established, and the AEM predictions (see Eqs.~(\ref{SSS})-(\ref{equ:M5})) can be verified directly.

\section{Results and discussion}
\subsection{Scaling laws of $\tau_{x,o}^{'+}$}
\begin{figure} 
	\centering 
	\subfigure{ 
		\label{fig:SSS:a} 
		\includegraphics[width=2.5in]{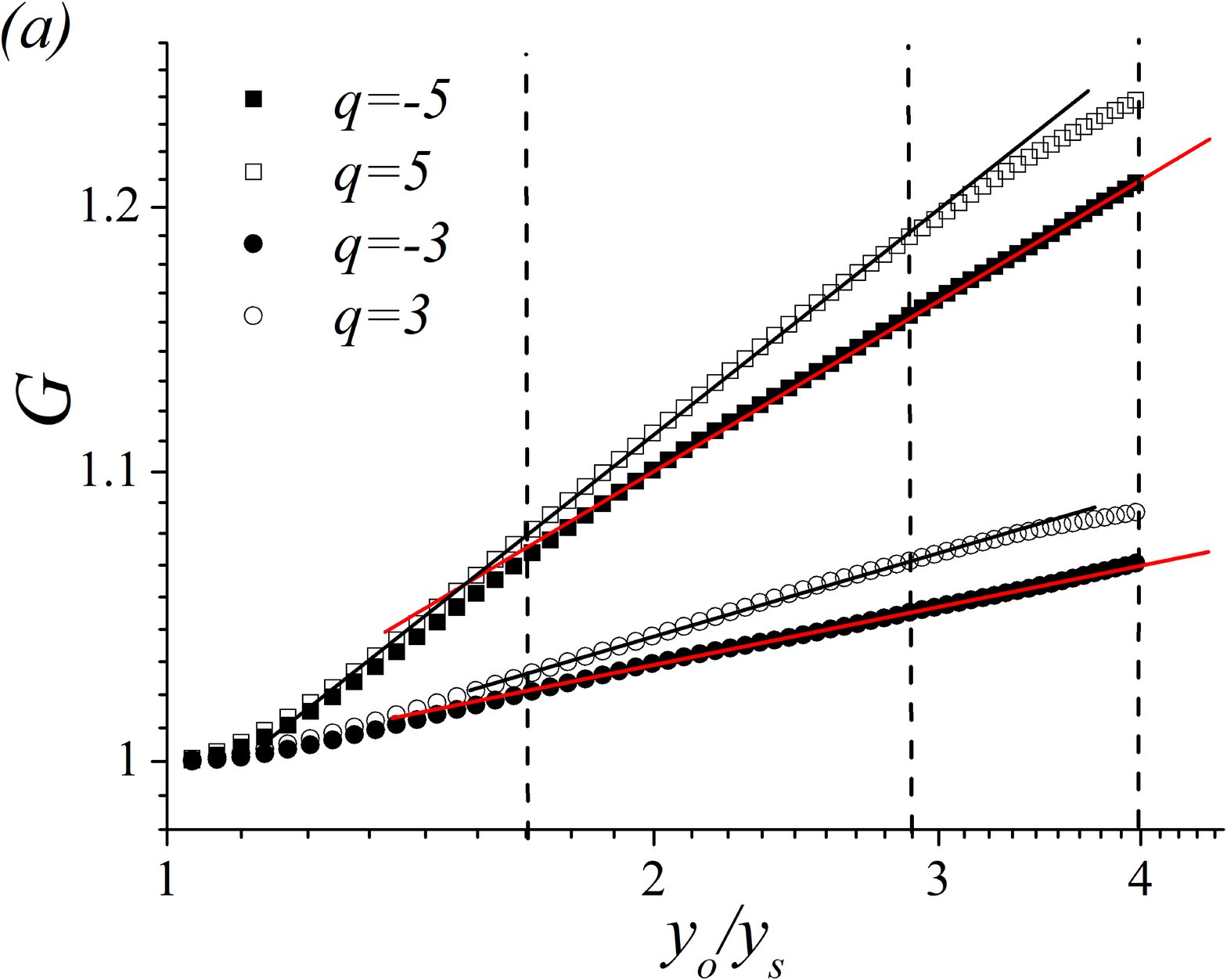} 
	}
	\subfigure{ 
			\label{fig:SSS:b} 
			\includegraphics[width=2.5in]{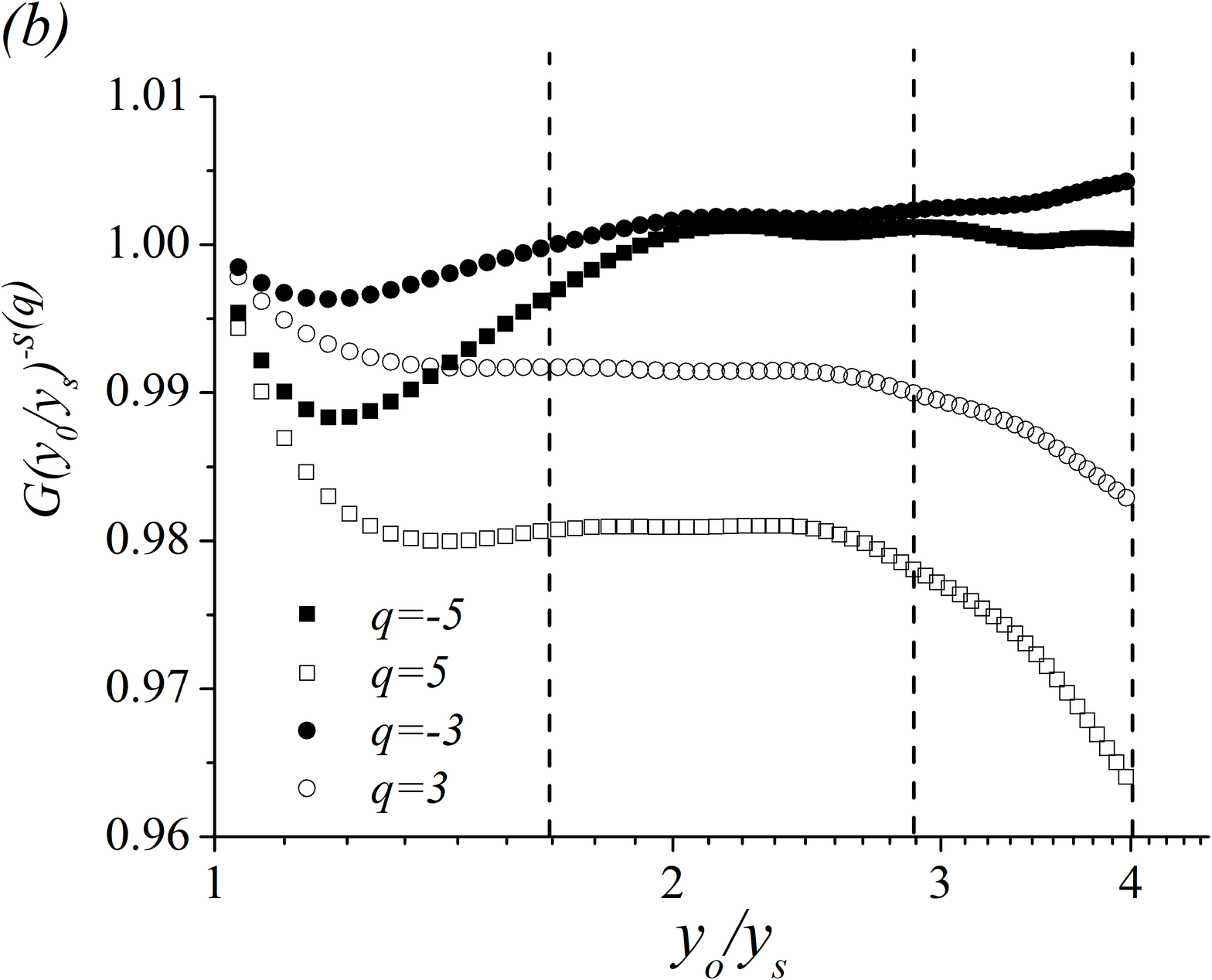} 
		}
	\caption{($a$) $G$ as functions of $y_o/y_s$ for $q=\pm5$ and $q=\pm3$; ($b$) premultiplied $G$ as functions of $y_o/y_s$ for $q=\pm5$ and $q=\pm3$.}
	\label{fig:SSS} 
\end{figure}
Here, we further define a moment generation function based on the IOIM, i.e.,
\begin{equation}
G(q,y_o^+)=\left\langle exp(q(\tau_{x,L}^{'+}(y_s^+)-\tau_{x,L}^{'+}(y_o^+)))\right\rangle .
\end{equation}
Fig.~\ref{fig:SSS}($a$) shows the variations of $G$ as a function of $y_o/y_s$ for $q=\pm5$ and $q=\pm3$. Power-law behaviours can be found in the interval between $1.7\le y_o/y_s\le2.9$ for positive $q$ and $1.7\le y_o/y_s\le4$ for negative $q$, justifying the validity of SSS, i.e., Eq.~(\ref{SSS}). Fig.~\ref{fig:SSS:b} is in aid  of accessing the scalings by displaying the variations of  premultiplied $G$. This observation highlights that the  superpositions of wall-attached log-region motions on wall surface follow the additive process, characterized by Eq.~(\ref{Gaa}). It is also worth mentioning that the power-law behaviour can be observed for larger wall-normal intervals for negative $q$. As $G(q,y_o^+)$ quantifies $\tau_{x,L}^{'+}(y_s^+)-\tau_{x,L}^{'+}(y_o^+)$, which features the same sign as $q$, this observation is consistent with the work of  \cite{Cheng2020a}, which showed that the footprints of the inactive part of attached eddies populating the logarithmic region are  actively connected with large-scale negative $\tau'_x$. Other $q$ values yield similar results and are not shown here for brevity.

\begin{figure} 
	\centering 
	\subfigure{ 
			\label{fig:SSS2:a} 
			\includegraphics[width=2.5in]{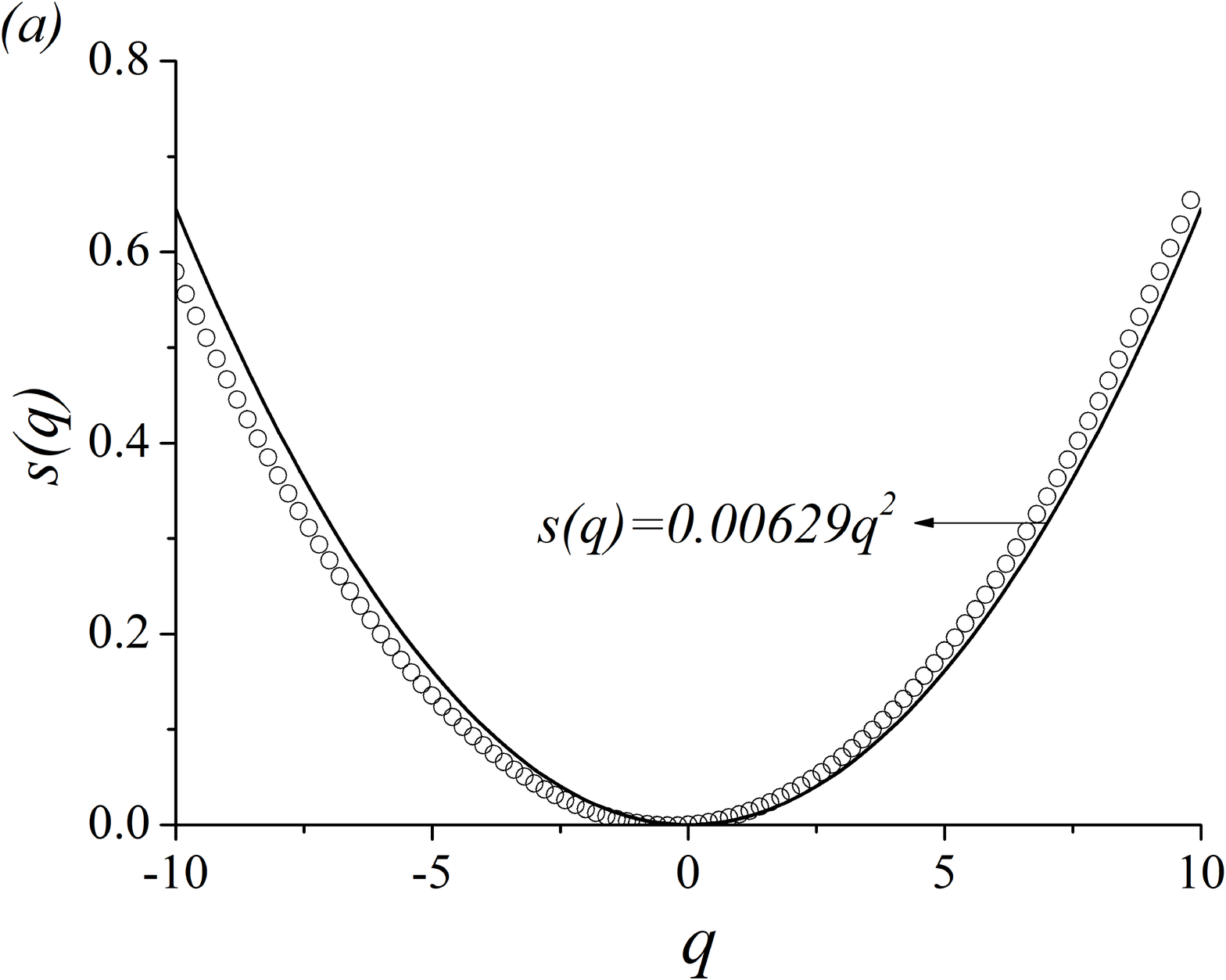} 
		}
	\subfigure{ 
				\label{fig:SSS2:b} 
				\includegraphics[width=2.5in]{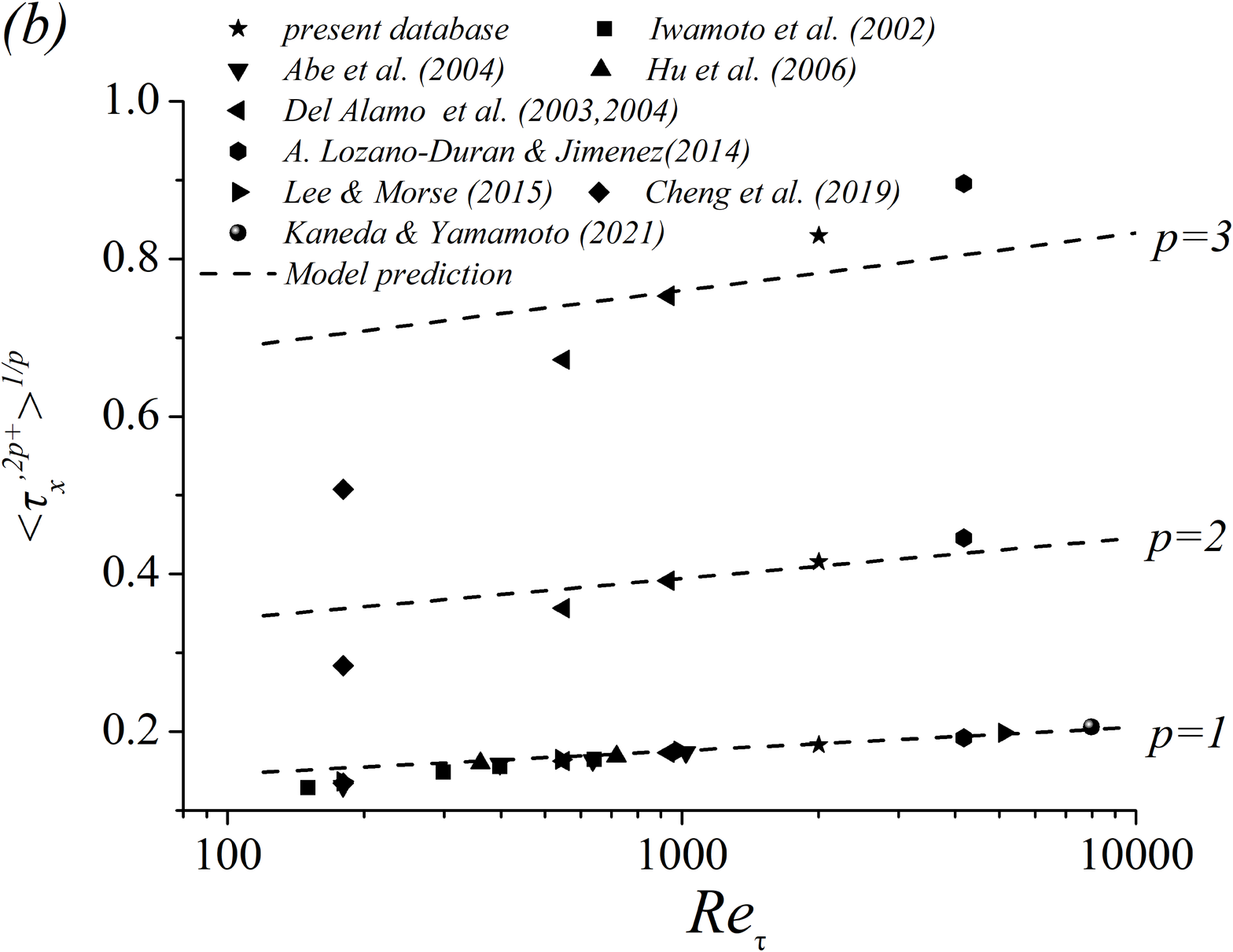} 
			}
	\caption{($a$) Anomalous exponent $s(q)$ as a function of $q$. The black line is a quadratic fit; ($b$) second- to sixth- order  moments of $\tau_{x}^{'+}$ as  functions of $Re_{\tau}$. The dashed lines are the log-normal predictions from Eq.~(\ref{Ga2})-(\ref{Ga6}).}
	\label{fig:SSS2} 
\end{figure}

The anomalous exponent $s(q)$ can be obtained by fitting the range $2\le y_o/y_s\le2.9$, where both positive and negative $q$ display good power-law scalings. Fig.~\ref{fig:SSS2}($a$) displays the variation of the anomalous exponent $s(q)$ as a function of $q$. The solid line denotes the quadratic fit within $-0.5\le q \le 0.5$. It can be seen that the variation of $s(q)$ is very close to the model prediction, i.e., the quadratic function as Eq.~(\ref{Ga}) with $C_2=0.00629$. Only minor discrepancies between DNS data and model predictions can be observed. As such, it is reasonable to hypothesize that the streamwise wall-shear stress fluctuation $\tau'_x$ generated by attached eddies of a given size follows the Gaussian distribution. Moreover, we can also estimate the statistical moments of $\tau_{x,o}^{+}$ by taking the derivative of $G(q,y_e^+)$ with respect to $q$ around $q=0$ \citep{Yang2016a}, i.e.,
\begin{equation}\label{Ga2}
\left\langle \tau_{x,o}^{'2+} \right\rangle =\left.\frac{\partial^{2} G(q ;y_o^+)}{\partial q^{2}}\right|_{q=0} \sim 2C_2\ln(y_o/y_s) \sim 2C_2\ln Re_{\tau},
\end{equation}
\begin{equation}\label{Ga4}
\left\langle \tau_{x,o}^{'4+} \right\rangle^{1/2}=(\left.\frac{\partial^{4} G(q ;y_o^+)}{\partial q^{4}}\right|_{q=0})^{1/2} \sim 2\sqrt{3}C_2\ln(y_o/y_s) \sim 2\sqrt{3}C_2\ln Re_{\tau},
\end{equation}
\begin{equation}\label{Ga6}
\left\langle \tau_{x,o}^{'6+} \right\rangle^{1/3} =(\left.\frac{\partial^{6} G(q ;y_o^+)}{\partial q^{6}}\right|_{q=0})^{1/3} \sim 2\sqrt[3]{15}C_2\ln(y_o/y_s) \sim 2\sqrt[3]{15}C_2\ln Re_{\tau}.
\end{equation}
Fig.~\ref{fig:SSS2}($b$) shows the variations of second- ($p=1$) to sixth- ($p=3$) order  moments of $\tau'_x$ calculated from DNS of channel flows \citep{Iwamoto2002,DelAlamo2003,Abe2004,DelAlamo2004,Hu2006,Lozano-Duran2014a,Lee2015,Cheng2019,Kaneda2021} and compares them with the model prediction, i.e., Eq.~(\ref{Ga2})-(\ref{Ga6}). For the second- and fourth- order variances, the model predictions are roughly consistent with the DNS results. The comparisons also indicate a Reynolds-number dependence of $\left\langle \tau_{x}^{'2+} \right\rangle$, which has been reported by vast studies \citep{Schlatter2010,Mathis2013,Guerrero2020}, and may be ascribed to the superposition effects of the wall-attached log-region motions. \cite{Wang2020} speculated that the amplitude modulation effect  plays a more prominent role in affecting the statistic characteristics of $\tau_{x,rms}^{'+}$ than the superposition effect, which contradicts the present findings. In fact, amplitude modulation has been demonstrated to exert a negligible effect on the even-order moments \citep{Mathis2011,Blackman2019}. Therefore, the deduction of  \cite{Wang2020} needs to be revisited. For sixth-order moments, the model prediction  displays substantial discrepancies with the DNS data. It is expected since high-order moments are dominated by the rare events resulting from the intermittent small-scale motions \citep{Frisch}, which can not be captured by IOIM (see Fig.~\ref{fig:pdf}($a$)).

\begin{figure} 
	\centering 
	\subfigure{ 
			\label{fig:ESS:a} 
			\includegraphics[width=2.5in]{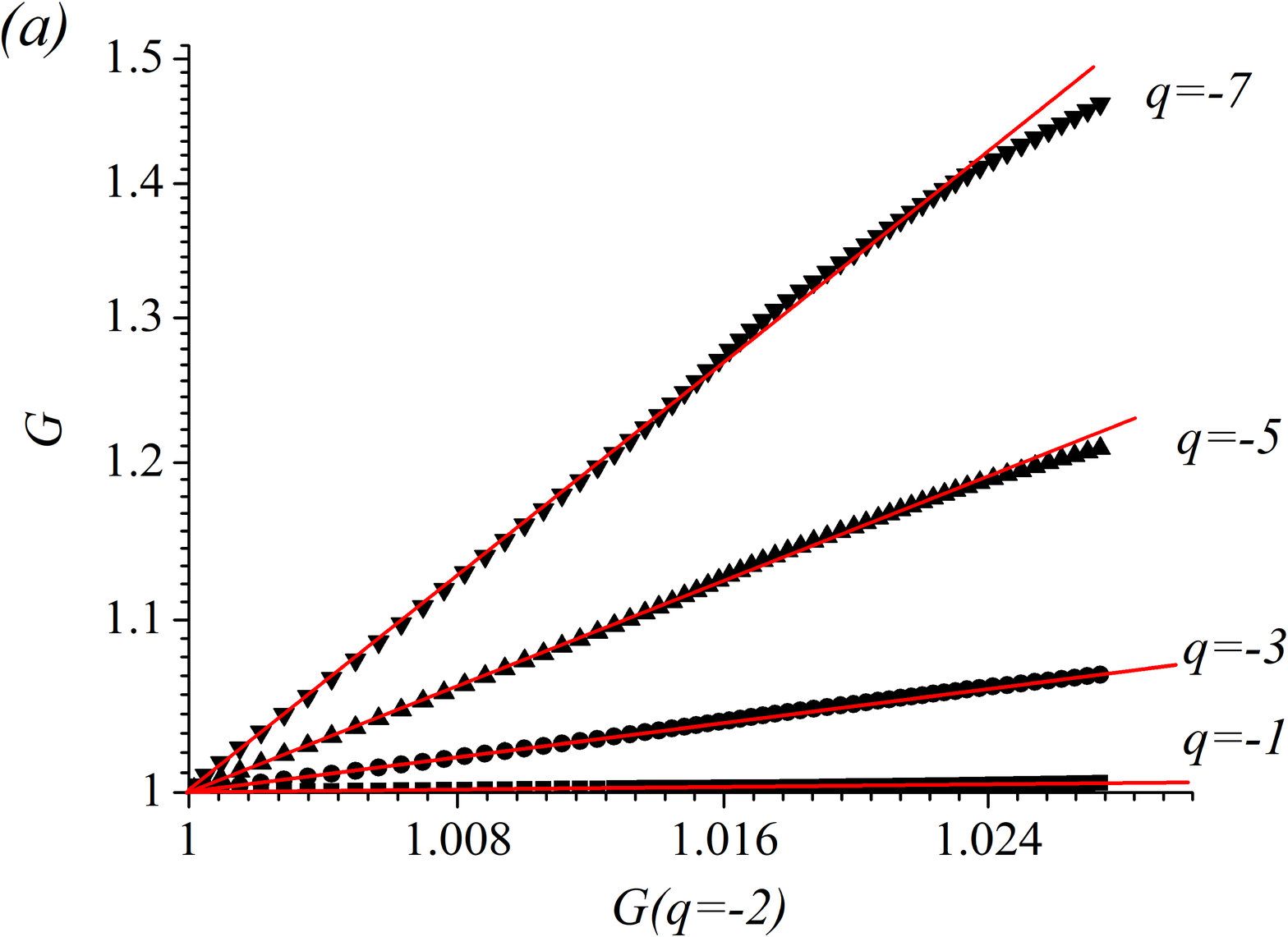} 
		}
	\subfigure{ 
				\label{fig:ESS:b} 
				\includegraphics[width=2.5in]{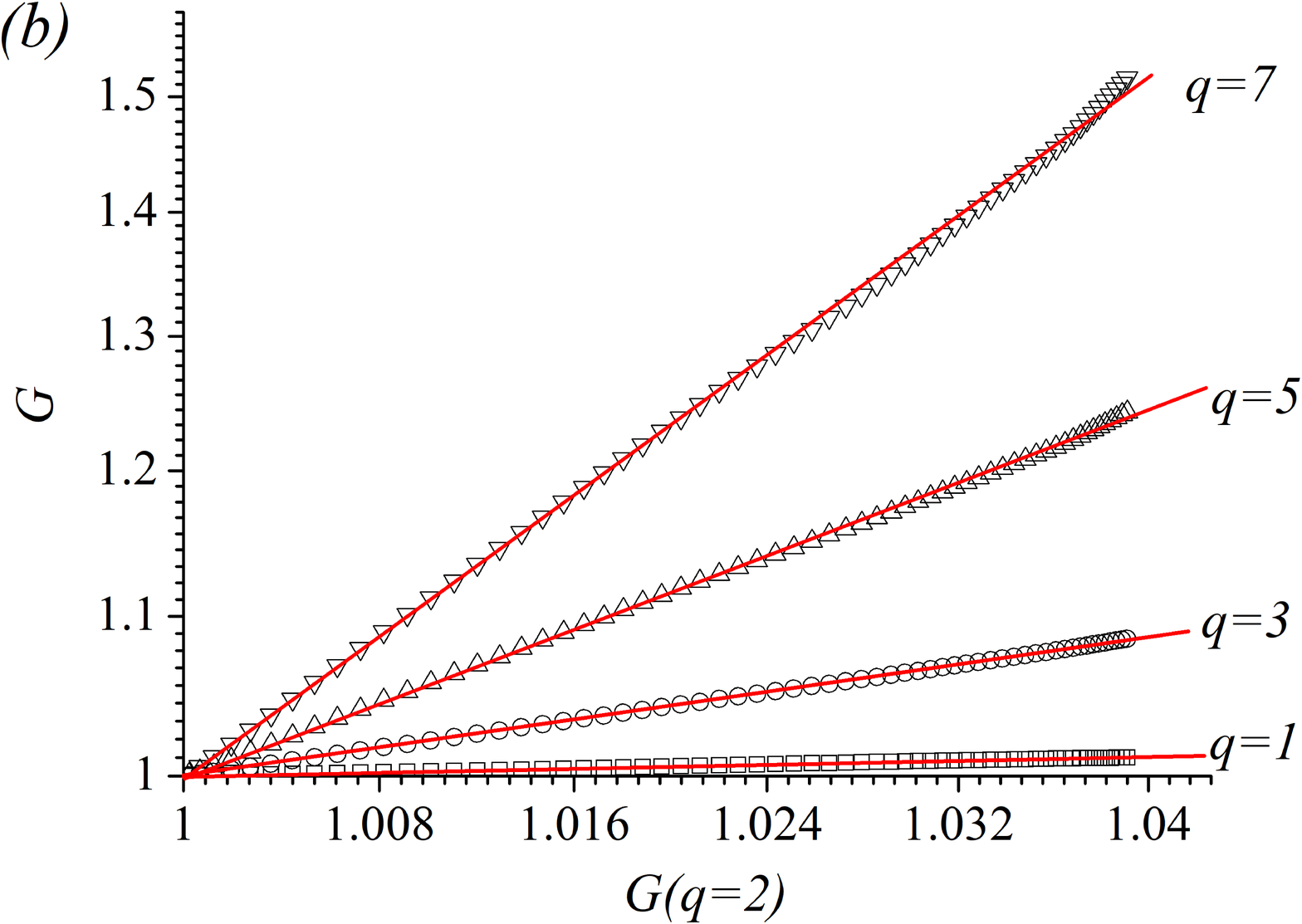} 
			}
	\caption{($a$) $G(q)$ as functions of $G(-2)$ for $q=-1,-3,-5,-7$; ($b$) $G(q)$ as functions of $G(2)$ for $q=1,3,5,7$. Both vertical and horizontal axes in ($a$) and ($b$) are plotted in logarithmic form.}
	\label{fig:ESS} 
\end{figure}

ESS (i.e., Eq.~(\ref{equ:M5})) is another scaling predicted by the multifractal formalism. Different from SSS, ESS does not rely on the i.i.d of the addends, but the additive process (see Eq.~(\ref{Gaa})).
Fig.~\ref{fig:ESS}($a$) and \ref{fig:ESS}($b$) shows the ESS scalings for $q_0=-2$ and $q_0=2$, respectively. ESS holds for the entire logarithmic region. The observation suggests that the streamwise wall-shear fluctuations generated by logarithmic motions obey the additive process, though the streamwise wall shear fluctuations generated by attached eddies with wall-normal heights at approximately $0.2h^+$ are not identically and independently distributed due to the scale interactions (see Fig.~\ref{fig:SSS}), which are not described by the attached-eddy model.

\subsection{Instantaneous distribution of $\tau'_x$}
\begin{figure} 
	\centering 
	\subfigure{ 
			\label{fig:PDF:a} 
			\includegraphics[width=2.5in]{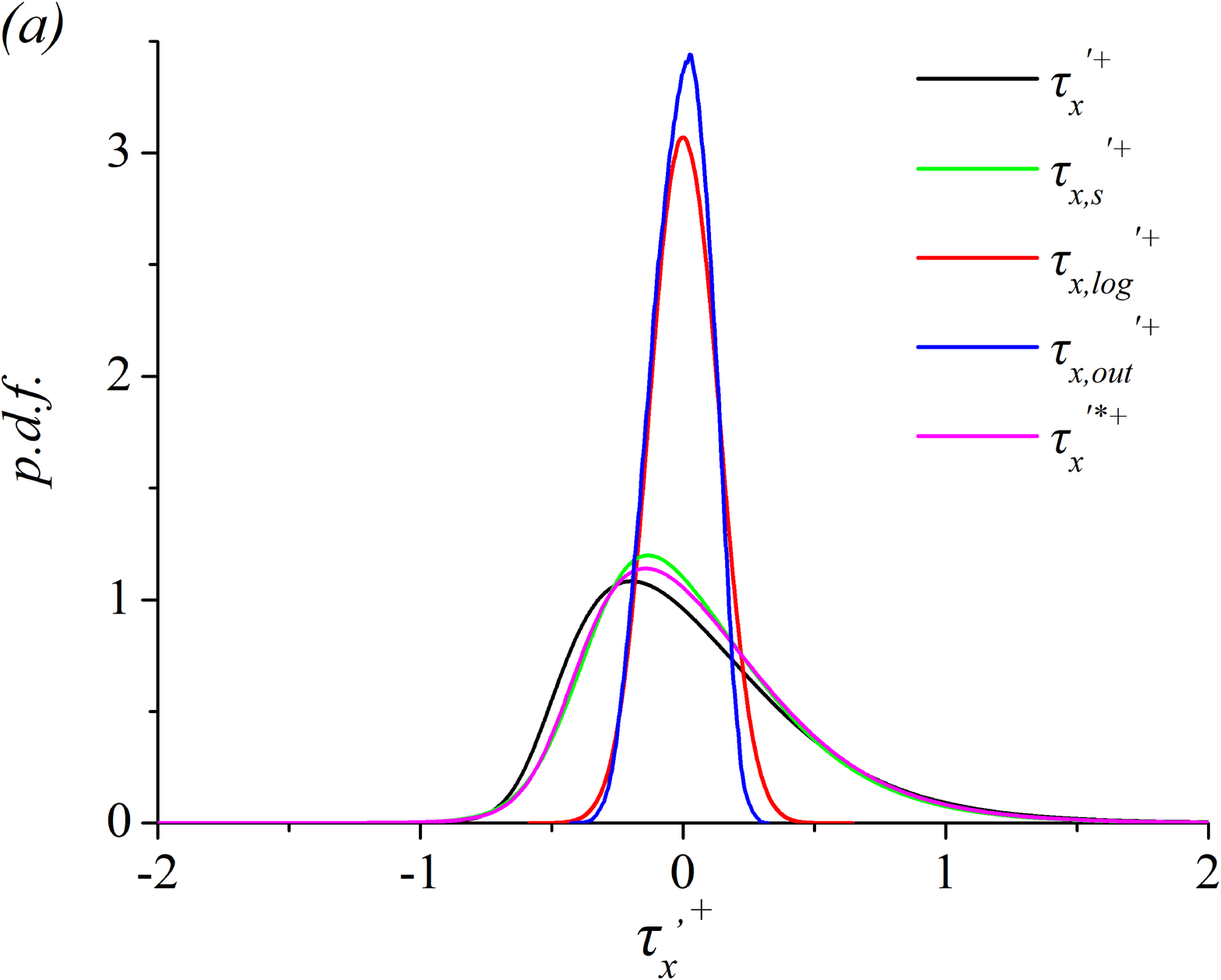} 
		}
	\subfigure{ 
				\label{fig:PDF:b} 
				\includegraphics[width=2.5in]{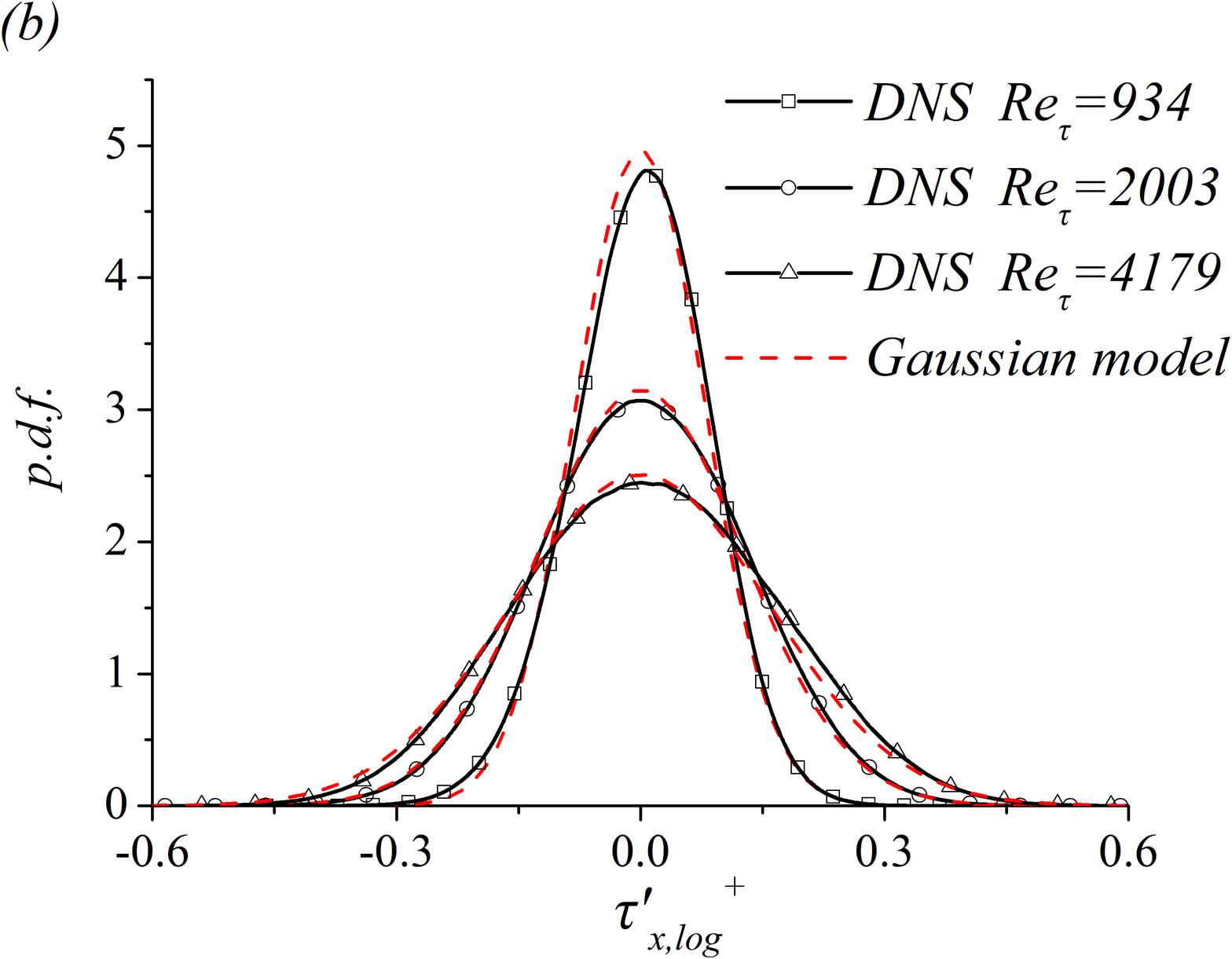} 
			}
	\caption{($a$) P.d.f.s of $\tau_{x}^{'*+}$, $\tau_{x,s}^{'+}$, $\tau_{x,log}^{'+}$,  $\tau_{x,out}^{'+}$, and $\tau_{x}^{'+}$; ($b$) P.d.f.s of  $\tau_{x,log}^{'+}$ in channel flows with $Re_{\tau}=934$, $2003$, and $4179$. Dashed lines denote the Gaussian model predictions with Eqs.~(\ref{pp1})-(\ref{pp2}).}
	\label{fig:pdf} 
\end{figure}

Furthermore, the instantaneous $\tau_x^{'+}$ can be decomposed as
\begin{equation}\label{MIOP2}
\tau_x^{'+}=\tau_{x,s}^{'+}+\underbrace{\tau_{x,L}^{'+}(y_s^+)-\tau_{x,L}^{'+}(y_e^+)}_{\tau_{x,log}^{'+}}+\underbrace{\tau_{x,L}^{'+}(y_e^+)}_{\tau_{x,out}^{'+}},
\end{equation}
where $\tau_{x,s}^{'+}$ denotes the amplitude modulation of the universal signal $\tau_{x}^{'*+}$, 
$\tau_{x,log}^{'+}$ and $\tau_{x,out}^{'+}$ are the superposition components contributed from the log region and the outer wall-coherent motions, respectively. The methodology of removing modulation effects can be found in \cite{Mathis2011} and \cite{Baars2016}, whose details are out of the range of present study. Fig.~\ref{fig:pdf}($a$) shows the probability density functions (p.d.f.s) of $\tau_{x}^{'*+}$, $\tau_{x,s}^{'+}$, $\tau_{x,log}^{'+}$ and $\tau_{x,out}^{'+}$, and compares with the p.d.f for the full channel data. The p.d.f.s of $\tau_{x,s}^{'+}$ and $\tau_{x}^{'*+}$  nearly coincide with that of $\tau_{x}^{'+}$ with asymmetric and positively skewed shape, which demonstrates that removing the superposition and modulation effects barely affects the instantaneous distributions. The asymmetries between the positive and negative wall-shear fluctuations are the essential characters of the near-wall small-scale turbulence, which may be associated with the celebrated near-wall sustaining process \citep{Schoppa2002}.
In contrast, the p.d.f.s of $\tau_{x,log}^{'+}$ and  $\tau_{x,out}^{'+}$ are more symmetric with rare events invisible, suggesting that the superposition components of logarithmic and outer motions are less  intermittent than the small-scale universal signals. This also explains the reason why the log-normal model describes the additive process well (see Fig.~\ref{fig:SSS2}($a$)), although the log-normal model is inapplicable for rare events \citep{Landau}. Moreover, the skewness and flatness of $\tau_{x,log}^{'+}$  are  0.05 and 2.91, which are very close to those of a Gaussian distribution. It strongly supports the conclusion drawn above that the  streamwise wall-shear stress fluctuations generated by attached eddies populating logarithmic region can be absolutely treated as Gaussian variables with
\begin{equation}\label{pp1}
p(\xi)=\frac{1}{\sqrt{2 \pi} \sigma} \exp \left(-\frac{\xi^{2}}{2 \sigma^{2}}\right),
\end{equation} 
where $p(\xi)$ denotes the p.d.f, and $\xi$ is the independent variable. It is worth noting that the variation of variance can be well predicted by the log-normal model, namely
\begin{equation}\label{pp2}
\sigma^{2}=\left.\frac{\partial^{2} G(q ;y_o^+)}{\partial q^{2}}\right|_{q=0}=2C_2\ln(Re_{\tau})+C_3,
\end{equation}
where $C_2\approx0.00629$, and $C_3\approx-0.07959$ is a constant and determined by the DNS data at $Re_{\tau}=2003$. Fig.~\ref{fig:pdf}($b$) shows the p.d.f.s of $\tau_{x,log}^{'+}$ and the model
prediction by Eq.~(\ref{pp1}), results of other two Reynolds numbers \citep{DelAlamo2004,Lozano-Duran2014a} are also included for comparison. It can be seen that the Gaussian model proposed here works reasonably well and can cover a wide range of Reynolds numbers. The model remains to be validated by higher-Reynolds number DNS data.
%

 \section{Concluding remarks}
In summary, the present study reveals that IOIM and AEM are consistent to each other quantitatively. The statistical characteristics of the superpositions of log-region eddies follow the predictions of AEM, namely, the SSS and ESS scalings. Based on these observations, we conclude that the streamwise wall-shear stress fluctuations generated by attached eddies populating the logarithmic region can be treated as Gaussian variables. 
A Gaussian model is then proposed to describe their instantaneous distributions  and verified by DNS data spanned broad-band Reynolds numbers. Considering the fact that the intensity of wall-shear stress fluctuations is typically underpredicted by the state-of-the-art wall-modelled large-eddy simulation (WMLES) approaches \citep{Park2016}, the Gaussian model proposed in the present study may be constructive for the development of the LES methodology, and the distribution characteristics of $\tau_{x}^{'*+}$ are helpful for developing more accurate near-wall models of WMLES approaches.

It is noted that some previous works adopted IOIM to investigate the spectral characteristics of the wall-coherent components of the signals in near-wall region, such as the work of \cite{Marusic2017}, but whether they are consistent with the AEM predictions in physical space quantitatively has not been verified in detail. The consistency of the two models demonstrated here fills the gap and complements their works. Moreover, the findings in present study indicate that we can isolate the footprints of attached eddies within a selected wall-normal range by employing IOIM, i.e., by adjusting $y_1^+$ and $y_2^+$ in $\tau_{x,L}^{'+}(y_1^+)-\tau_{x,L}^{'+}(y_2^+)$. Here $y_1^+$ and $y_2^+$ are two selected wall-normal heights in the logarithmic region, and $y_1^+<y_2^+$. In this regard, the present study may provide a new perspective for analyzing some flow physics in wall-bounded turbulence, such as the inner peak of the intensity of $u'$, and the streamwise inclined angles of attached eddies. All these are under investigation currently and will be reported in separate forthcoming papers.

\section*{Acknowledgments}
 We are grateful to the authors cited in Fig.~\ref{fig:SSS2}($b$) for making their invaluable data available. L.F. acknowledges the fund from CORE as a joint research center for ocean research between QNLM and HKUST.

\section*{Declaration of interests}
The authors report no conflict of interest.

\bibliographystyle{jfm}
\bibliography{eddy2}

\end{document}